\documentclass[12pt,epsf]{article}
\usepackage{graphicx, amsmath}

\begin{document}

\sloppy
\begin{flushright}{SIT-HEP/TM-43}
\end{flushright}
\vskip 1.5 truecm
\centerline{\large{\bf Running spectral index from shooting-star moduli}}
\vskip .75 truecm
\centerline{\bf Tomohiro Matsuda\footnote{matsuda@sit.ac.jp}}
\vskip .4 truecm
\centerline {\it Laboratory of Physics, Saitama Institute of Technology,}
\centerline {\it Fusaiji, Okabe-machi, Saitama 369-0293, 
Japan}
\vskip 1. truecm
\makeatletter
\@addtoreset{equation}{section}
\def\theequation{\thesection.\arabic{equation}}
\makeatother
\vskip 1. truecm

\begin{abstract}
\hspace*{\parindent}
We construct an inflationary model
that is consistent with both large non-Gaussianity and a running
spectral index. 
The scenario of modulated inflation suggests that modulated perturbation
can induce the curvature perturbation with a large non-Gaussianity, even
 if the inflaton perturbation is negligible.
Using this idea, we consider a multi-field extension of the modulated inflation scenario
and examine the specific situation where different moduli are
 responsible for the perturbation at different scales.
We suppose that the additional moduli (shooting-star moduli) is
 responsible for the curvature perturbation at the earlier inflationary
 epoch and it generates the fluctuation with $n>1$ spectral index at
 this scale.
After a while, another moduli (or inflaton) takes the
 place and generates the perturbation with $n<1$. 
At the transition point the two fluctuations are comparable with each
 other. 
We show how the spectral index is affected by the transition induced by
 the  shooting-star moduli. 
\end{abstract}

\newpage
\section{Shooting-Star moduli}
\hspace*{\parindent}
String theory and supersymmetric models generically predict 
many flat directions and moduli that determine the coupling constants in
the effective action.
Assuming that at least a few of these fields are light during inflation, 
the vacuum fluctuations of the light scalar fields
${\cal M}_i$ are unstable and appear as classical random Gaussian
inhomogeneities with an almost scale-free spectrum of amplitude
$\delta {\cal M}_i\simeq H_I/2\pi$, where $H_I$ is the Hubble parameter
during inflation.
Then the wavelength of the fluctuations is stretched during inflation
over the Hubble horizon after inflation. 
This is the reason why we believe in this paper that the ``modulated
fluctuations'' 
$\delta {\cal M}_i$ can be related in many different ways to the
cosmological curvature perturbation in the present Universe.  
Let us first review the basic idea of modulated inflation that has been
discussed by us in Ref.\cite{modulated-inflation}.
Our starting point is the conventional equation for the number of
e-foldings elapsed during inflation;
\begin{equation}
\label{original_1}
N =\frac{1}{M_p^2}\int^{\phi_N}_{\phi_e} \frac{V}{V_\phi}d\phi,
\end{equation}
where $\phi_N$ is the value of inflaton field $\phi$ corresponding to
$N$ e-foldings, and $\phi_e$ denotes the end-point of inflation where
the slow-roll condition is violated.
Using $\delta N$-formula we can see that the fluctuation of a common
spectrum $\delta \phi_N = H_I/2\pi$ induces the spectrum of the density
perturbation given by 
\begin{equation}
 \delta_H^2 = \frac{4}{25}(\delta N)^2 
=\frac{4}{25}\left(\frac{V}{M_p^2V_\phi}\frac{H_I}{2\pi}\right)^2,
\end{equation}
where we followed the notations given in the textbook\cite{Lyth-book}.
Of course, in more generic situation one may expect several scalar
fields (moduli or flat directions) that may play a similar role during
inflation. 
The first specific example in this direction has been given by
Bernardeau et 
al.\cite{modulated_hybrid} for modulated couplings in hybrid-type
inflation, assuming that $\phi_e$ depends on such a light field.
Note that the word ``modulated fluctuations'' was introduced
by Kofman in Ref.\cite{Kof-1}.
Then Lyth\cite{At_the_end_of} considered a multi-inflaton model of
hybrid inflation and encounters another realization of 
``generating the curvature perturbation at the end of inflation''.
And more recently, we considered trapping inflation combined with
inhomogeneous preheating\footnote{Note that thermal inflation is induced
by ``thermal trapping'', while trapping inflation is induced by
``trapping by the preheat field''.} and found a different
mechanism for generating the curvature perturbation at the end of
inflation\cite{at_the_end_of_trapped}. 
The multi-inflaton models\cite{At_the_end_of,
at_the_end_of_trapped} are 
very useful for brane inflation, since there can be several directions
for the fluctuation of the moving brane, as well as of the target brane.
See Ref.\cite{brane-deltaN} for more details of the fluctuation related
to the target brane.
In fact, the multi-inflaton model has been applied to brane  
inflation to solve the serious $\eta$-problem in
string theory\cite{brane-deltaN}. 
Note that in string theory it is very hard to find a light field that
is perfectly suitable for the conventional single-field inflation,
especially when the potential of moduli fields 
are determined by some calculable mechanism of moduli stabilization.
It is therefore very helpful if a light field other than the
inflaton can contribute to the curvature perturbation. 
In fact, there are many alternatives for the conventional scenario,
in which the inflaton fluctuation $\delta \phi\ne 0$ plays no role
in generating the cosmological fluctuation.
In those ``alternatives'', light fields other than the inflaton
play crucial roles in generating
the cosmological perturbation.
For example, cosmological perturbation can be generated (1) long after
inflation (curvatons)\cite{curvaton_1,curvaton_low, hilltop-curvaton},
(2) during 
preheating (inhomogeneous preheating)\cite{inhomogeneous_preteating}, or
(3) during reheating (inhomogeneous
reheating)\cite{inhomogeneous_reteating}, and also by combining (1) and
(2) one can generate the initial perturbation of the curvaton from
inhomogeneous preheating\cite{curvaton_from_PR}.
Note that large non-Gaussianity can be generated during inhomogeneous
preheating, and moreover, even if the non-Gaussianity is not generated
by the  inhomogeneous preheating, small ratio $r\equiv
\rho_{\chi}/\rho_{total} \ll 1$ at the decay can 
lead to a large non-Gaussianity $f_{nl}\propto r^{-1}$.
This mechanism is similar to the one that has been discussed for 
curvatons\cite{curvaton-fnl}.\footnote{See also
Ref.\cite{fnl_from_PR_other}.}
Here $\rho_{\chi}$ denotes the energy density of the preheat field.
As a result, $f_{nl}$ in inhomogeneous preheating scenarios can be large
and can take either (positive or negative) sign.
One of the reasons that we consider such ``alternatives'' is
that a large non-Gaussianity of the spectrum may be confirmed by the
observation\cite{large-non}. 

The idea of modulated inflation in
Ref.\cite{modulated-inflation} is very simple.
Let us look at the equation (\ref{original_1}).
Besides the inflation fluctuations related to $\delta \phi_N$ and
$\delta \phi_e$, 
fluctuations induced by other components may 
generate curvature perturbation if these 
components are modulated during inflation.
Based on this simple idea, we considered an alternative
mechanism for generating the curvature
perturbation\cite{modulated-inflation}, which relies 
neither on $\delta \phi_N$ nor $\delta \phi_e$.
This distinguishes modulated inflation from the previous scenario of 
modulated fluctuation\cite{modulated_hybrid}. 

Besides the large non-Gaussianity that may exclude conventional
single-field inflation, there is another problem related to the running
of the spectral index.
The Wilkinson Microwave Anisotropy Prove(WMAP) data favor primordial
cosmological fluctuation with a spectral index $n>1$
at large scale and $n<1$ at smaller scale.
One way to generate the fluctuation with the required running 
spectrum is to
consider different forms of inflationary potential at different scales,
 and then merge them at the scale where $n$ passes through unity.
This possibility of running spectral index has been discussed by many
authors\cite{running-single} for conventional inflationary scenario.
However, these models may be excluded by
a large non-Gaussianity parameter\cite{large-non}, since conventional
inflationary scenario typically generates Gaussian perturbation. 

Therefore, our motivation in this paper is to construct a first concrete
example
that is consistent with both large non-Gaussianity and a running
spectral index. 
In this paper, we consider hybrid inflation with a simple 
moduli-dependent inflaton mass 
\begin{equation}
m^2({\cal M}_i)\equiv m^2_0\left(1+
\beta_1\frac{{\cal M}_1^2}{M_*^2}+
\beta_2\frac{{\cal M}_2^2}{M_*^2}\right),
\end{equation}
which induces fluctuation related to $\delta V_\phi$.\footnote{Note that
these moduli fields are not the inflaton in hybrid-type 
inflation, since they cannot lead to the waterfall.
In this respect, we are not considering a multi-field extension of the
hybrid-type inflation model. Moreover, the inflaton fluctuation is not
important in the modulated inflation scenario. Modulated inflation can
be discriminated from the multi-inflaton model by these
characteristics.}
For simplicity, we consider a specific case in which the inflaton
fluctuation is negligible.\footnote{Of course, the inflaton fluctuation
can collaborate with ``shooting-star'' moduli. 
We will consider this possibility in appendix A.}
In this specific example, ${\cal M}_2$ is the ``shooting-star'' moduli 
that has the positive $\eta$-parameter $\eta_2>0$ and is responsible for
the running of the spectral index at a larger scale.
We assume that the potential of the moduli fields are dominated by
simple quadratic term
\begin{equation}
V({\cal M}_i)=\frac{\eta_i H_I^2 {\cal M}_i^2}{2}.
\end{equation}
Then, from the equation of motion we find that during inflation the
value of the field ${\cal M}_2$ decreases as 
\begin{equation}
{\cal M}_2 \propto e^{-\eta_2\Delta N},
\end{equation}
where $\Delta N$ is the number of e-foldings elapsed during the
evolution.
We consider a parameter space where the effective mass
that may appear from the interaction with the inflaton is not
important.\footnote{See Appendix B for more details.}
At the earlier stage of inflaton, where the perturbation at the larger
scale is generated, the fluctuation of the
shooting-star moduli (i.e., $\delta {\cal M}_2$) generates
the dominant part of the cosmological fluctuation which has the spectral
index $n \simeq 2\eta_2 >1$.
Note that the derivative of $N$ with respect to ${\cal M}_2$ is given by
\begin{equation}
N_{2} \simeq 2N\beta_2 \left(\frac{{\cal M}_2}{M_*^2}\right),
\end{equation}
which {\bf decreases} with time.
On the other hand, we suppose that ${\cal M}_1$ has a $\eta$-parameter
$\eta_1<0$ and {\bf increases} slowly during inflation.
As a result, the fluctuation induced by the moduli ${\cal M}_1$ 
has the spectral index $n \simeq 2\eta_1<1$.
The derivative of $N$ with respect to ${\cal M}_1$ is given by
\begin{equation}
N_{1} \simeq 2N \beta_1\left(\frac{{\cal M}_1}{M_*^2}\right),
\end{equation}
which {\bf increases} with time.
Here we introduce dimensionless parameters 
\begin{eqnarray}
r_1 &\equiv& \frac{N_1^2}{N_1^2+N_2^2}\nonumber\\
r_2 &\equiv& \frac{N_2^2}{N_1^2+N_2^2},
\end{eqnarray}
and consider the spectral index\cite{Lyth-book}
\begin{eqnarray}
\label{index-multi}
n-1 &=& -\left(\frac{M_p V_{a}}{V}\right)^2
-\frac{2}{M_p^2 N_{a}^2}
+2 \frac{M_p^2 N_{a} N_{b} V_{ab}}{V N_{d} N_{d}}\nonumber\\
&\simeq & 2\eta_1 r_1
+ 2\eta _2 r_2,
\end{eqnarray}
where the subscript of the potential means the derivative with respect
to the corresponding field.
Terms related to the $\epsilon$-parameter is discarded because they
are small in this case.
Looking at Eq.(\ref{index-multi}), we find that at the transition from
the ${\cal M}_2$-dominated perturbation to the ${\cal M}_1$-dominated
one, there is a ``jump'' in the spectral index.
Note that the transition is a natural consequence of our set-up, and 
the time of the transition depends on the initial condition.
The running of the spectral index is given by
\begin{equation}
\frac{dn}{d\ln k} \simeq -4\eta_2^2r_2
\end{equation}
for $|\eta_2| \gg |\eta_1|$ and $r_1 \sim r_2$, which can be applied to
the data at the scale $k=0.002 Mpc^{-1}$.

Let us apply our results to the data.
Since we are considering a model in which the perturbation at the
smaller scale is generated by the 
${\cal M}_1$-perturbation, we find $\eta_1 <1$ and $r_2 \ll
1$ at that smaller scale.
Here we can use the WMAP3 data\cite{WMAP3} at the scale $k=0.05
Mpc^{-1}$;  
\begin{eqnarray}
\label{index-5}
n_{0.05} = 0.948^{+0.014}_{-0.018} &,& \frac{dn}{d\ln k} \sim 0.
\end{eqnarray} 

On the other hand, the value of the spectral index and its running 
at the larger scale $k=0.002 Mpc^{-1}$ is given by
\begin{eqnarray}
\label{data-wmap}
n = 1.21^{+0.13}_{-0.16} &,& \frac{dn}{d\ln k} = -0.102^{+0.050}_{-0.043},
\end{eqnarray}
which leads to the condition 
\begin{equation}
\eta_2 \sim 0.25
\end{equation}
and $r_2(k) \sim 0.4$ at $k=0.002 Mpc^{-1}$.
Of course, there are several ambiguities in this naive calculation.
In addition to the ambiguities in the above data for the value of the 
spectral index and the running, the 
ambiguity may also arise in the value of the scale parameter $k$.
In any case, we may conclude that the spectral index and the running
at the scale $k=0.002 Mpc^{-1}$ can be generated by the shooting-star
moduli, with the cost of tuning parameters and the 
initial condition.\footnote{Although the $\delta N$ formula that relates
the final curvature perturbation on comoving slices to the inflaton
perturbation on flat slices after horizon crossing is a very powerful
tool for our computation, there are at least two possibilities that must
be examined carefully. One is the inhomogeneous
reheating\cite{inhomogeneous_reteating} that may arise due to the
modulated decay rate ($\delta \Gamma$) of the inflaton field, and the
other is the possibility of generating the curvature perturbation at the
end of inflation. Of course, the latter is a part of the $\delta N$
formula, however the effect may not be obvious in a naive calculation.
We added Appendix B and C to discuss more precise conditions for our
results.}
Note also that the jump in the spectral index may occur several times
during inflation, if there are many shooting-star moduli(or flat
directions) in the theory.

\section{Conclusions and discussions}
\hspace*{\parindent}
We have studied a new class of modulated inflation that generates
the running of the spectral index at a larger scale.
We have shown a concrete example of modulated inflation in which 
the ``shooting-star'' moduli generates the running of the spectral index
at a larger scale.
As far as we know, the present model is the first and simple
concrete example of the inflation scenario that is consistent with both
the large non-Gaussianity and the running spectral index.
Perhaps one can construct a model in which a similar
transition occurs for curvatons, inhomogeneous
preheating and inhomogeneous reheating scenarios.
In fact, in these alternative scenarios the spectral index can be related
to the $\eta$-parameter of the light field\cite{hilltop-curvaton}.
An obvious deficit of the modulated scenario may be the famous moduli
problem. 
Late-time entropy production such as thermal inflation\cite{Lyth-book}
 may solve this
problem, but thermal inflation may not work if the energy scale of the 
primordial inflation is very low.
On the other hand, if the cosmological perturbation in modulated
inflation is due to a flat direction of a supersymmetric gauge theory,
the flat direction (i.e, ${\cal M}_i$) can decay 
fast through preheating.
In this case, the moduli problem may not occur.
Note that the multi-field models such as Ref.\cite{At_the_end_of,
at_the_end_of_trapped, inhomogeneous_preteating} are free from the
moduli problem because the light direction gets large mass soon after
the inflation.
Of course, it is always very hard to construct inflation model
that works with a low inflationary scale\cite{curvaton_low,
hilltop-curvaton, low-infla} despite the fact that low-scale inflation
may become important if the gravitational effect is observed at the
Large Hadron Collider(LHC).

\section{Acknowledgment}
We wish to thank K.Shima for encouragement, and our colleagues at
Tokyo University for their kind hospitality.
\appendix

\section{Shooting-star moduli that helps conventional inflation}

The shooting-star moduli may induce a transition from the
moduli-dominated perturbation to the inflaton-dominated one, -and vice
versa.
Therefore, the idea of the shooting-star moduli 
may be useful in explaining the running of the spectral index in
conventional inflation model.

For example, let us consider a case in which the inflaton perturbation
is responsible for the curvature perturbation at the scale smaller than
$k=0.05 Mpc^{-1}$.
Then we can find the condition for the inflaton potential as
usual\cite{Lyth-book}, using the data (\ref{index-5}) at that scale.
Note that unlike the conventional inflationary scenario, 
one does not have to worry about the running of the spectral index at a
larger scale, since the shooting-star moduli is responsible for the
running. 
The shooting-star moduli disappears from the spectrum soon after the
transition and is not observed at the smaller scale.
As a concrete example, we consider chaotic inflation with a quartic
potential 
\begin{equation}
V(\phi)=\frac{\lambda}{4}\phi^4
\end{equation}
and the moduli-dependent Planck mass
\begin{equation}
M_p({\cal M})=M_p\left(1+\beta \frac{{\cal M}^2}{M_*^2}\right).
\end{equation}
The $\delta N$-parameter related to the inflaton is
\begin{equation}
\delta N_\phi =\frac{\phi}{4M_p^2} \delta \phi,
\end{equation}
and the one related to the moduli is\cite{modulated-inflation}
\begin{equation}
\delta N_{\cal M}=-4N\beta \left(\frac{{\cal M}}{M_*^2}\right) 
\delta {\cal M}.
\end{equation}
At the smaller scale we need the spectral index $n<1$, which is
realized by $\delta N_\phi$ that dominates the perturbation at that scale.
On the other hand, at a larger scale the moduli fluctuation dominates the
perturbation and generates the spectral index $n>1$.
Therefore, these two perturbations are comparable at the transition.
The condition $\delta N_\phi \simeq \delta N_{\cal M}$ at that scale
leads to
\begin{equation}
\label{appA1}
|{\cal M}| \simeq \frac{\phi M_*^2}{16N \beta M_p^2}
\frac{\delta \phi}{\delta {\cal M}}
\simeq M_p \times
 \frac{\alpha_*^2\alpha_\delta}{4\sqrt{2N}\beta },
\end{equation}
where the definition of the dimensionless parameters are
$\alpha_* \equiv M_*/M_p$ and $\alpha_\delta 
\equiv \frac{\delta \phi}{\delta {\cal M}}$.
If a large non-gaussianity is generated by the moduli perturbation,
we find
\begin{equation}
f_{nl}\sim -\frac{1}{4N\beta}\frac{M_*^2}{{\cal M}^2}\simeq 
-\frac{\beta}{2\alpha_*^2\alpha_\delta^2}.
\end{equation}
As we have explained in this paper, the running of the spectral index 
can be generated if the 
shooting-star moduli has large\footnote{$\eta_{\cal M}$ is large
conpared with $\eta_\phi$, but it is smaller
than unity.} and positive $\eta$-parameter. 
The running of the non-Gaussianity is significant in this specific
example.

One may suspect that the modulated Planck mass may lead to the
generation of the curvature perturbation at the end of inflation.
In fact, $\phi_e$ is defined by the slow-roll parameter
$\epsilon(\phi_e) \simeq 1$, which leads to $\phi_e\simeq 4 M_p$.
Therefore, $M_p$ in this equation is determined by the value of the
moduli ${\cal M}_e$ at the end of inflation.
However, as far as we are considering the shooting-star moduli that
rolls down toward the origin during inflation, the value of the moduli
and its fluctuation at the end of inflation is significantly smaller
than the one that appeared in the above calculation.
We thus conclude that the perturbation generated at the end of inflation
is negligible in this specific example.

\section{Other conditions}

In this appendix, we consider other conditions that are needed for
realizing large non-Gaussianity and the running spectral index for the
hybrid-inflation model.
These conditions are highly model-dependent, but we hope they are
helpful for the analysis in this direction.

First, we consider the condition for the primordial
curvature perturbation generated by the modulated inflation;
\begin{eqnarray}
\label{appB1}
\delta N &\simeq& N_1 \delta {\cal M}_1\nonumber\\
&\simeq& 2N \beta_1  \frac{{\cal M}_1}{M_*^2} \frac{H_I}{2\pi} 
\simeq 5\times 10^{-5}.
\end{eqnarray}
If a large non-Gaussianity is generated by the ${\cal M}_1$ moduli,
there is a condition 
\begin{eqnarray}
\label{appB2}
-\frac{3}{5} f_{nl} &\simeq& \frac{1}{2}
\frac{\partial^2 N/\partial {\cal M}_1^2}
{\left(\partial N/\partial {\cal M}_1\right)^2}\nonumber\\
&\simeq& \frac{M_*^2}{4N\beta_1 {\cal M}_1^2}.
\end{eqnarray}
From Eq.(\ref{appB1}) and (\ref{appB2}), we find
\begin{equation}
\label{appBa}
{\cal M}_1 \simeq 10^{4}\frac{H_I}{f_{nl}}.
\end{equation}

The effective mass that is generated by the interaction between the
moduli and the inflaton field must be small.
Here we put the condition
\begin{equation}
|\beta_i| \frac{m_0^2 \phi^2}{M_*^2} \le |\eta_i| H_I^2,
\end{equation}
which leads to the bound
\begin{equation}
\label{appB3}
|\phi| \le \frac{M_*}{\sqrt{|\beta _i|}}.
\end{equation}
Our assumption in this paper is that the perturbation generated 
by the conventional inflaton fluctuation does not dominate the
cosmological perturbation.
We thus need the condition $N_\phi < N_1 $, which leads to
\begin{equation}
\frac{1}{|\eta_\phi \phi|}< 2N|\beta_1| \frac{{\cal M}_1}{M_*^2}.
\end{equation}
Note that this condition gives the lower bound for the inflaton;
\begin{equation}
\label{appB4}
|\phi| > \frac{M_*^2}{2N |\beta_1\eta_\phi| {\cal M}_1}.
\end{equation}
Combining Eq.(\ref{appB3}) and (\ref{appB4}), we find
\begin{equation}
\label{appBb}
{\cal M}_1 > \frac{M_*}{|\eta_\phi| N \sqrt{|\beta_1|}}.
\end{equation}
Since ${\cal M}_1<M_*$ is a natural condition in this model,
we find
\begin{equation}
|\eta_\phi| N \sqrt{|\beta_1|} >1
\end{equation}
We find that fast-roll inflation is favored in this specific
example.\footnote{Fast-roll hybrid inflation is
discussed by Dimopoulos et.al\cite{fast-roll}.}
Note that fast-roll inflaton field with the $\eta$-parameter larger than
unity does not generate the primordial perturbation.
Therefore, the condition (\ref{appB4}) does not appear if $\eta_\phi
\ge1$. 
Finally, from Eq.(\ref{appBa}) and (\ref{appBb}), we find the lower
bound for the inflation scale; 
\begin{equation}
H_I > \frac{f_{nl}M_*}{10^4 |\eta_{\phi}| N \sqrt{|\beta_1|}}.
\end{equation}

One may also think that the modulated couplings may induce the 
generation of the cosmological perturbation at the end of inflation.
However, at least in the present example, the fluctuations of the moduli
fields do not induce significant fluctuation in $\delta \phi_e$,
since the moduli fields do not appear in the equation that determines
$\phi_e$. 
In this respect, modulated inflation can be discriminated from
multi-inflaton model in which the curvature perturbation generated 
(converted) near the end of inflation is crucial.

\section{Inhomogeneous reheating}
One may suspect that the modulated couplings that have been used in this
paper may lead to inhomogeneous (modulated) reheating after inflation.
If so, a significant perturbation may be generated at the 
reheating, which may ruin the model.

Let us first take a look at the hybrid inflation model.
The reheating temperature of a hybrid inflation model is
determined by the effective mass near the true minimum.
Therefore, at least in the hybrid inflation model that has been
discussed in this paper, the model is free from such problem, since
there is no modulated fluctuation in the inflaton decay rate that
remains until reheating. 

On the other hand, in the chaotic inflation model that we considered in
Appendix A, the fluctuation of the Planck mass may induce inhomogeneous
reheating after inflation. 
If this effect is larger than the preceding perturbation, the running of
the spectral index and a larger non-Gaussianity may be erased at the
reheating.
This can be a serious problem in our scenario.
We thus need to explain the conditions that are needed to avoid
the problem in the chaotic inflation model.
Since the problem may arise when inflaton decays through gravity-mediated
interaction, we consider a decay rate that is proportional to
$M_p^{-n}$, which leads to the fluctuation
\begin{equation}
\frac{\delta \Gamma}{\Gamma} \simeq 
-n \frac{\partial M_p/\partial {\cal M}}{M_p}
\delta {\cal M}
\simeq -2n\beta \frac{{\cal M}}{M_*^2} \delta {\cal M},
\end{equation}
where the values of the fields are evaluated at the reheating.
Since we are considering a shooting-star moduli, ${\cal M}$
decreases rapidly during inflation.
Therefore, the expectation value of the moduli ${\cal M}$ is 
much smaller than the one that has been used for the calculation of the
running spectral index.
Our conclusion is that at least in the specific example that was
considered in Appendix A, inhomogeneous reheating does not lead to 
the generation of a significant perturbation after inflation.

\end{document}